\documentclass[12pt]{article}
\usepackage{amsfonts}
\usepackage{amssymb}
\usepackage{amscd}

\begin{document}

\centerline{\bf ON UNCERTAINTY RELATIONS AT PLANCK'S SCALE}

\bigskip
\bigskip
\bigskip
\centerline{\bf Branko Dragovich }
\smallskip
\bigskip

\centerline{ Institute of Physics, P.O. Box 57, 11001 Belgrade,
Yugoslavia}

\centerline{Steklov Mathematical Institute, 117966 Moscow, Russia}

\centerline{ E-mail: dragovich@phy.bg.ac.yu}

%\maketitle
%\begin{abstract}
\bigskip
\bigskip
\bigskip

\noindent{\bf Abstract.} A brief review of  the previous research
on the Heisenberg uncertainty relations at the Planck scale is
given. In this work, investigation of the uncertainty principle
extends to $p$-adic and adelic quantum mechanics. In particular,
$p$-adic analogs of the Heisenberg algebra and uncertainty
relation are introduced. Unlike ordinary quantum theory, adelic
quantum approach provides a promising framework to probe space
below the Planck length.

\bigskip
\bigskip
\bigskip

\noindent{\bf 1. Introduction}

\bigskip

\noindent At macroscopic scales spacetime is a pseudo-Riemannian
manifold locally identified as the Minkowski space. Position $x$
and momentum  $k$ of a particle are described by classical
mechanics, and precision of their measurements is not formally
restricted. However, at microscopic scales the Heisenberg
uncertainty relation emerges: $$ \Delta x \, \Delta k \geq
\frac{\hbar}{2} \,,  \eqno(1.1) $$ which is one of the basic
characteristics of quantum mechanics. According to (1.1), position
and momentum  of the same particle cannot be simultaneously well
determined. Relation (1.1) is not dynamical, but kinematical in
its origin, and it is  a realization of the associative
noncommutative Heisenberg algebra:
$$ [\hat{x},\hat{k}] =i\,\hbar \,.  \eqno(1.2) $$
However, the combination of quantum and general relativity
principles, which becomes significant at the Planck scale
$$\ell_0
= \sqrt{\frac{\hbar \, G}{c^3}} \sim 10^{-33} \mbox{cm} \,,
\eqno(1.3)
$$ leads, by means of gedanken experiments, to the modification of
relation (1.1) (for some recent references, see [1] - [3]). So,
several studies (see, e.g. [4,5]) in string theory, which is the
best candidate for unification of all interactions (including
gravitational force), yield
$$ \Delta x \, \Delta k \geq
\frac{\hbar}{2} \left(1 + a\, \frac{\lambda^2}{\hbar^2}\, (\Delta
k)^2\right) \,, \quad a>0 \,,  \eqno(1.4) $$ where $\lambda$ is
the string length $\ell_s$ and $a$ is a constant. In some other
quantum gravity considerations [6] $\lambda$ is the Planck length
$\ell_0$ (1.3). Since the relation (1.4) is a quadratic algebraic
inequality, it implies
$$\Delta x \geq \sqrt{a}\, \lambda \,,\eqno(1.5) $$
i.e. there exists a finite minimal spatial uncertainty $\Delta x_0
= \sqrt{a} \,\lambda$. It is usually adopted $\Delta x_0 =
\ell_0$.  Relation (1.4) should be regarded as a consequence of
the fact that a string as a probe is a nonlocal object and extends
itself with increasing energy.  Formally (1.4) can be obtained
from the algebra
$$ [\hat{x},\hat{k}]= i\, \hbar \, (1 + \alpha \, \hat{k}^2),
\ \ \  \alpha = a \, \lambda^2/\hbar^2 \,,  \eqno(1.6)$$ which is
a modification of (1.2).

In addition to the modification (1.4) of the standard
position-momentum uncertainty relation (1.1), there is also an
investigation within string theory which results in  spacetime
uncertainty relation  (see, e.g. [7])
$$ \Delta T\, \Delta X \geq \ell_s^2 \,. \eqno(1.7)  $$
A generalization of (1.7) to the uncertainty relation for the
worldvolume of a Dq-brane was also proposed [8]:
$$ \Delta X^0\, \Delta X^1 \cdots \Delta X^q \geq g_s \, \ell_s^{q+1} ,
\eqno(1.8) $$ where $g_s$ is a string coupling.

On the basis of (1.7), (1.8), and some other considerations of
noncommutativity (for a review, see [9]), there is a sense to
introduce spatial noncommutativity in the form
$$ [\hat{x}^m, \hat{x}^n ] = i\, \theta \, \varepsilon^{mn}  \,, \quad \varepsilon^{mn}
 = - \varepsilon^{nm} \,,  \eqno(1.9) $$
what gives $\Delta x^m \, \Delta x^n \geq \theta/2 \,$ when $m\neq
n$. Since $\Delta x \geq \ell_0$ it follows that $\theta \geq
2\ell_0^2$. It means that spatial uncertainty at Planck's scale is
influenced by both (1.6) and (1.9) noncommutativity.

Note that existence of the minimal uncertainty $\Delta x \geq
\ell_0$ means that one cannot experimentally investigate pieces of
space smaller than the Planck length. In this way, standard
approach to the Planck scale physics based on real (and complex)
numbers only, predicts it own breakdown at distances smaller than
$\ell_0$. This is one of the main reasons to introduce into the play
$p$-adic numbers and adeles.
\bigskip
\bigskip

\noindent{\bf 2.  $p$-Adic Uncertainty Relations}
\bigskip

\noindent $p$-Adic numbers [10] and adeles [11] have been
effectively applied in some physical models since 1987 (for a
review, see [12] and [13]). Recall that any $p$-adic number $x
\neq 0$ can be uniquely presented in the form
$$
x = p^{-\nu}\, (x_0 + x_1 \, p + x_2 \, p^2 + \cdots) \,, \quad
\nu \in {\mathbb Z}\, ,\,\, x_0 \neq 0\,, \,\, x_i \in \{0, 1,
2,\cdots , p-1 \} \,,  \eqno(2.1)
$$
and its $p$-adic norm is $|x|_p = p^\nu$. If $\nu \geq 1$ then
rational part of $x$ is $\{ x \}_p = p^{-\nu} \, (x_0 + x_1 p +
\cdots + x_{\nu -1} p^{\nu-1} )$, but if $\nu \leq 0$ then $\{
x\}_p = 0$. The field of rational numbers ${\mathbb Q}$ is dense
subfield in ${\mathbb R}$ as well as in ${\mathbb Q}_p$ for any
prime number $p$.  (All other necessary properties of $p$-adic
numbers and related analysis can be found in Refs. [10]-[13].)

Here we investigate $p$-adic uncertainty relations in the
framework of $p$-adic [14] and  adelic [15] quantum mechanics . An
adelic eigenfunction has the form
$$
\psi (x) = \psi_\infty (x_\infty) \prod_{p\in S} \psi (x_p)
\prod_{p\notin S} \Omega (|x_p|_p) \,,  \eqno(2.2)
$$
where $S$ is a finite set of primes $p$, and $\Omega (|x_p|_p) =
1$ if $|x_p|_p \leq 1$, and $\Omega (|x_p|_p)=0$ if $|x_p|_p > 1$.
In the eq. (2.2), $\psi_\infty (x_\infty)$ and $\psi_p (x_p)$ are
eigenfunctions of ordinary and $p$-adic quantum mechanics,
respectively, and $\Omega (|x_p|_p)$ is the simplest $p$-adic
vacuum state.

For an observable ${\mathcal D}_p$ we define the corresponding
$p$-adic expectation value $<{\mathcal D}_p>$ and uncertainty
$\Delta {\mathcal D}_p$ as generalization of these properties in
conventional quantum mechanics. Namely,
$$
<{\mathcal D}_v> = \int_{{\mathbb Q}_v} \psi^\ast_v (x)\,
{\mathcal D}_v \, \psi_v (x) \, dx \,, \quad \Delta {\mathcal D}_v
= \big( <{\mathcal D}_v^2> - < {\mathcal D}_v>^2
\big)^{\frac{1}{2}} \,, \eqno(2.3)
$$
where $v=\infty$ and $v=p$ are related to standard and $p$-adic
cases, respectively. Since coordinate $x_p$ is $p$-adic valued,
and $\psi_p$ as well as $\Omega$-function are complex-valued, it
is not possible to calculate expectation value $<x_p>$ and
uncertainty $\Delta x_p$ of $x_p$, but one can do that for
$p$-adic norm  $|x_p|_p$  and rational part $\{ x_p\}_p$ of $x_p$
, as well as for the corresponding momenta $k_p$ . Here we
consider the following cases: ${\mathcal D}_p = |x_p|_p\,\, , \{
x_p\}_p \,\,, |k_p|_p \,\,, \{k_p\}_p \, $. For simplicity, we
will often write $x$ and $k$ instead of $x_p$ and $k_p$. As
illustrative $p$-adic quantum states $\psi_p (x)$ we take three
simple and typical (vacuum) eigenfunctions. Recall that the
Fourier transform of $\psi_p (x)$ is  $\tilde{\psi}_p (k) =
\int_{{\mathbb Q}_p} \chi_p (k x)\, \psi_p (x)\, dx \,$.

According to (2.2), it is of particular interest to start with the
case $(i)$:   $\, \, \psi_p (x) =\Omega (|x|_p) \,, \quad \psi_p
(k) =\Omega (|k|_p) $ with
$$
<|x|_p>\, =\,  <|k|_p>\, = \,  \int_{|x|_p \leq 1} |x|_p \, dx =
\frac{1-p^{-1}}{1-p^{-2}} \rightarrow 1 \,, \, \, p\rightarrow
\infty \,, \eqno(2.4)
$$
$$
\Delta |x|_p \, = \, \Delta |k|_p = \left[
\frac{1-p^{-1}}{1-p^{-3}} - \left( \frac{1-p^{-1}}{1-p^{-2}}
\right)^2 \right]^{\frac{1}{2}} \rightarrow 0\,, \,\, p
\rightarrow \infty \,, \eqno(2.5)
$$
$$
<\{ x\}_p> \, =\, <\{ k\}_p>\, = \, \int_{|x|_p \leq 1} \{ x\}_p
\,\, dx = 0 \,, \quad \Delta \{x\}_p = \Delta \{k \}_p = 0 \, .
\eqno(2.6)
$$

As the second example we use orthonormal eigenstates $(ii)$
$\psi_p (x) = p^{\frac{\nu}{2}}\, \Omega (p^\nu \, |x|_p)\,,\, \,
\, \tilde{\psi}_p (k) = p^{-\frac{\nu}{2}}\, \Omega (p^{-\nu} \,
|k|_p)\,, \,\,\, \nu \in {\mathbb Z}$. For this class of $p$-adic
quantum states, we have:
$$ <|x|_p> = p^{-\nu}\,\frac{1-p^{-1}}{1-p^{-2}} ,\ \ \ \ <|k|_p>
= p^\nu \, \frac{1-p^{-1}}{1-p^{-2}} \,,  \eqno(2.7)
$$
$$
\Delta |x|_p = p^{-\nu}\, \left[ \frac{1-p^{-1}}{1-p^{-3}} -
\left( \frac{1-p^{-1}}{1-p^{-2}}  \right)^2 \,
\right]^{\frac{1}{2}} \,, \eqno(2.8)
$$
$$\ \  \Delta |k|_p = p^{\nu} \,
\left[ \frac{1-p^{-1}}{1-p^{-3}} - \left(
\frac{1-p^{-1}}{1-p^{-2}} \right)^2  \, \right]^{\frac{1}{2}}\,,
 \eqno(2.9)
$$
$$
<\{ x \}_p> =\left\{ \begin{array}{ll} \frac{p^{-\nu} -1}{2
p^{-\nu}} \,, \,\, & \nu = 0, -1 , -2,, \cdots \,, \\ 0 \,,  & \nu
= 1, 2, 3, \cdots \,, \end{array} \right.           \eqno(2.10)
$$
$$
<\{ k \}_p> =\left\{ \begin{array}{ll} 0 \,,  \, \,\, & \nu = 0,
-1 , -2,, \cdots \,, \\  \frac{p^{\nu} -1}{2 p^{\nu}}  \,,  & \nu
= 1, 2, 3, \cdots \,,
\end{array} \right.           \eqno(2.11)
$$
$$
\Delta \{ x\}_p = \left\{ \begin{array}{ll}  \frac{1}{2
p^{-\nu}}\, \sqrt{\frac{p^{- 2\nu} - 1}{3}} \,, \, \,\, & \nu =0,
-1, -2, \cdots \,, \\ 0 \,,   \, \,\, &  \nu = 1, 2, 3, \cdots \,,
\end{array}  \right.           \eqno(2.12)
$$
$$
\Delta \{ k\}_p = \left\{ \begin{array}{ll} 0  \,, \, \,\, & \nu
=0, -1, -2, \cdots \,, \\ \frac{1}{2 p^{\nu}}\, \sqrt{\frac{p^{
2\nu} - 1}{3}} \,,   \, \,\, &  \nu = 1, 2, 3, \cdots \, .
\end{array}  \right.           \eqno(2.13)
$$

As the last example, let us take eigenstates $( iii)$: $\psi_p(x)
= p^{-\frac{\nu}{2}}\, (1-p^{-1})^{-\frac{1}{2}} \, \delta (p^\nu
- |x|_p)\,, \, \quad  \tilde{\psi}_p(k) = p^{\frac{\nu}{2}} \,
(1-p^{-1})^{-\frac{1}{2}} \, [\Omega (p^\nu|k|_p) - p^{-1} \Omega
(p^{\nu -1} |k|_p) ] \,, \, \, \nu \in {\mathbb Z} \,, \, $  where
$\delta (p^\nu -|x|_p) =1 $ if $|x|_p =p^\nu$ and $\delta (p^\nu
-|x|_p) =0 $ if $|x|_p \neq p^\nu$. Performing relevant
calculations, one obtains:
$$
<|x|_p> = p^\nu, \ \  <|k|_p>=2\, p^{-\nu} \,
\frac{1-p^{-1}}{1-p^{-2}} \,,\eqno(2.14)
$$
$$
\Delta |x|_p =0, \ \ \Delta |k|_p = p^{-\nu} \, \left[
\frac{1+p-2\,p^{-1}}{1-p^{-3}} - \frac{4\,
(1-p^{-1})^2}{(1-p^{-2})^2} \right]^{\frac{1}{2}} \,, \eqno(2.15)
$$
$$
<\{ x \}_p> = \left\{ \begin{array}{ll} 0 \,, \quad & \nu = 0, -1,
-2, \cdots \,, \\ \frac{1}{2} \,, \quad & \nu = 1, 2, 3, \cdots
\,,   \end{array}   \right.       \eqno(2.16)
$$
$$
<\{ k \}_p> = \left\{ \begin{array}{lll} \frac{1 - p^{\nu -1} \,
(p-1)}{2} \,, \quad & \nu =  -1, -2, -3, \cdots \,, \\ \frac{1}{2
\, p} \,, \quad &  \nu = 0 \,, \\ 0 \,, \quad & \nu = 1, 2, 3,
\cdots \,,   \end{array}   \right.              \eqno(2.17)
$$
$$
\Delta \{ x\}_p = \left\{ \begin{array}{ll} 0\,, \quad & \nu = 0,
-1, -2, \cdots \,, \\   \frac{1}{2 \, p^\nu} \, \sqrt{\frac{p^{2\,
\nu -1} - 2}{3 \, p^{-1}}} \,, \quad  &\nu = 1, 2, 3, \cdots \,,
\end{array}   \right.    \eqno(2.18)
$$
$$
\Delta \{k \}_p  = \left\{ \begin{array}{lll}  \frac{1}{2} \,
\sqrt{\frac{1 - p^{2 \nu} + 4 p^{2 \nu -1} - 5 p^{2 \nu - 2}}{3}}
\,, \quad  &  \nu = -1, -2, -3, \cdots \,, \\ \frac{1}{2 \, p}\,
\sqrt{\frac{4 p - 5}{3}} \,, \quad & \nu = 0 \,, \\ 0 \,, \quad &
\nu = 1, 2, 3, \cdots \, . \end{array}  \right.     \eqno(2.19)
$$

It is worth noting that some the above wave functions $( i)-(
iii)$ are eigenstates of the four-dimensional  quantum
cosmological de Sitter model, as well as for some other models,
for which $\ell_0$ emerges naturally and it is taken to be $\ell_0
= 1$ [16].

\bigskip
\bigskip
\newpage
\noindent{\bf 3. Discussion and  Concluding Remarks}

\bigskip

\noindent As was stated in the Introduction there is a minimal
uncertainty  $\Delta x = \ell_0 = 1$ in the ordinary theory of
space-time described by real numbers. Consequently one cannot use
an ordinary quantum gravity, based on real and complex numbers
only, to probe space at distances smaller than the Planck length.
It is therefore natural to ask the following question: Is it in
principle possible to measure distances smaller than $\ell_0$
using adelic quantum approach instead of the conventional one? To
give an answer let us look at the above derived mean values and
related uncertainties. To this end, note that $|x|_p$ plays a role
of the $p$-adic distance between points $0$ and $x$, and
consequently $\Delta |x|_p$ is a possible uncertainty measuring
the distance $|x|_p$. Rational part $\{ x\}_p$ contains more
information on $x$ than $|x|_p$, and hence $<\{ x \}_p>$ may be
regarded as a $p$-adic analog of $<x>$ in the real case . The
above obtained results depend on the prime number $p$ and integer
$\nu$.

{\bf Case (i).} In the $\Omega (|x|_p)$ states, which are
ingredients of any adelic eigenfunction for all but a finite
number of $p$, we have $1> \,\, <|x|_p> \, \rightarrow 1$ as
$p\rightarrow \infty$ and $1 > \, \Delta |x|_p \rightarrow 0$ when
$p \rightarrow \infty$. We have also $<\{x \}_p> = \Delta \{ x\}_p
= 0$ for every $p$.

{\bf Case (ii).} One can choose such enough large positive $\nu$
to have arbitrary small $<|x|_p>$ and $\Delta |x|_p$. For any
nonnegative integer $\nu$ one has $<\{ x \}_p> = \Delta \{ x\}_p =
0$.

{\bf Case (iii).} For these states $\Delta |x|_p = 0$ and
$<|x|_p>$ can be made arbitrary small for suitable negative $\nu$.
If $\nu$ are not positive integers then we have $<\{ x\}_p> =
\Delta \{ x \}_p = 0$.

From the above consideration one can conclude that adelic quantum
theory allows to probe space beyond the Planck length, and it is
possible owing to $p$-adic quantum effects. Now we can introduce
$p$-adic commutation relations in an analogous way to ordinary
quantum mechanics.

According to the Weyl quantization let us introduce operators
(with $h = 1$) $\,\, X_\beta (\hat{x})$ and $K_\alpha (\hat{k})$,
which act on $\psi \in L_2 ({{\mathbb Q}_v})$ as follows:
$$
X_\beta (\hat{x}) \, \psi (x) = \chi_v (-\beta \, \hat{x})\,
\int_{{{\mathbb Q}_v}} \chi_v (- k\, x) \, \tilde{\psi} (k)\, dk =
\int_{{{\mathbb Q}_v}} \chi_v (- (k + \beta)\, x) \, \tilde{\psi}
(k)\, dk   \,,                \eqno(3.1)
$$
$$
K_\alpha (\tilde{k})\, \psi (x) = \chi_v (- \alpha \, \hat{k}) \,
\int_{{{\mathbb Q}_v}} \chi_v (- k\, x) \, \tilde{\psi} (k)\, dk
$$
$$
= \int_{{{\mathbb Q}_v}} \chi_v (- ( x + \alpha)\, k) \,
\tilde{\psi} (k)\, dk = \psi (x + \alpha)\, ,  \eqno(3.2)
$$
where $\chi_\infty (a) = \exp{(- 2 \pi i a)}$ and $\chi_p (a) =
\exp{(2 \pi i \{a \}_p)}$ are real and $p$-adic additive
characters, respectively.

As a result, we have commutation relation
$$
\chi_v (-\beta \, \hat{x}) \, \chi_v (-\alpha \, \hat{k}) = \chi_v
(\alpha\, \beta) \, \chi_v (- \alpha \hat{k}) \, \chi_v (-\beta \,
\hat{x}) \,,                          \eqno(3.3)
$$
which after an expansion of characters yields (1.2) in the real
case and
$$
\{ \beta \, \hat{x} \}_p \, \{ \alpha \, \hat{k} \}_p - \{ \alpha
\, \hat{k} \}_p \, \{ \beta \, \hat{x} \}_p = -\frac{i}{2 \pi}\,
\{ \alpha \, \beta \}_p \,,  \quad p = 2, 3, 5, \cdots
\eqno(3.4)
$$
in the $p$-adic one. Starting from
$$
\int_{{\mathbb Q}_p} |\, a\, \hat{A}_p \psi (x) + i\, \hat{B}_p \,
\psi (x)\, |_\infty^2 \, dx \geq 0 \,,              \eqno(3.5)
$$
where $\hat{A}_p = \{ \beta \, \hat{x}\}_p - <\{\beta \,x \}_p>
\,, \quad   \hat{B}_p = \{ \alpha \, \hat{k}\}_p - <\{\alpha \, k
\}_p> $ and $|\cdot |_\infty$ denotes usual absolute value, we
obtain
$$
\Delta \{ \beta \, x\}_p \, \Delta \{ \alpha \, k\}_p \geq
\frac{1}{4 \pi} \, \{ \alpha \, \beta \}_p \,, \quad h=1.
\eqno(3.6)
$$

It is worth noting that $p$-adic uncertainty relation (3.6)
contains a point $(\alpha\,, \beta)$ of classical phase space
around which we consider uncertainty of quantum quantities $x$ and
$k$. We see that uncertainty highly depends on this point. The
same relation, where index $p$ is replaced by $\infty$, can be
derived from ordinary quantum mechanics. Uncertainty relations
obtained in Sec. 2 may be regarded as particular cases of (3.6).

At the end, it is worth pointing out that a string as an adelic
object [17] with increasing energy extends in the real metric but
shrinks in the $p$-adic ultrametric. Thus ultrametricity opens
window to the space-time below Planck's scale.

\bigskip
\bigskip
\noindent{\bf Acknowledgements}. The author would like to thank
organizers of the 12th Yugoslav Conference on Nuclear and Particle
Physics (Studenica, September 2001) for creative scientific
atmosphere. This work is partially supported by Serbian research
project 01M01, as well as by RFFI grant 990100866.

\end{document}